\documentclass[twocolumn]{aastex63}

\received{}
\revised{}
\accepted{}

\submitjournal{The Astrophysical Journal}

\usepackage{rotating}
\usepackage{gensymb}

\shorttitle{Microinstabilities in the Shock Transition Region}
\shortauthors{Kim et al.}

\begin{document}

\title{Microinstabilities in the Transition Region of Weak Quasi-Perpendicular Intracluster Shocks}

\author[0000-0002-5441-8985]{Sunjung Kim}
\affil{Department of Physics, School of Natural Sciences UNIST, Ulsan 44919, Korea}
\author[0000-0001-7670-4897]{Ji-Hoon Ha}
\affiliation{Department of Physics, School of Natural Sciences UNIST, Ulsan 44919, Korea}
\author[0000-0002-5455-2957]{Dongsu Ryu}
\affiliation{Department of Physics, School of Natural Sciences UNIST, Ulsan 44919, Korea}
\author[0000-0002-4674-5687]{Hyesung Kang}
\affiliation{Department of Earth Sciences, Pusan National University, Busan 46241, Korea}
\correspondingauthor{Sunjung Kim}
\email{sunjungkim@unist.ac.kr}

\begin{abstract}


Microinstabilities play important roles in both entropy generation and particle acceleration in collisionless shocks. 
Recent studies have suggested that 
in the transition region of quasi-perpendicular ($Q_{\perp}$) shocks
in the high-beta ($\beta = P_{\rm gas}/P_{\rm B}$) intracluster medium (ICM),
the ion temperature anisotropy due to the reflected-gyrating ions could trigger the Alfv\'en ion cyclotron (AIC) instability and the ion-mirror instability,
while the electron temperature anisotropy induced by magnetic field compression could excite the whistler instability and the electron-mirror instability.
Adopting the numerical estimates for ion and electron temperature anisotropies found in the particle-in-cell (PIC) simulations of $Q_{\perp}$-shocks with sonic Mach numbers, $M_{\rm s} = 2-3$,
we carry out a linear stability analysis for these microinstabilities. 
The kinetic properties of the microinstabilities and the ensuing plasma waves on both ion and electron scales are 
described for wide ranges of parameters, including $\beta$ and the ion-to-electron mass ratio. 
In addition, the nonlinear evolution of the induced plasma waves are examined by performing 2D PIC simulations with periodic boundary conditions. 
We find that for $\beta\approx 20-100$, the AIC instability could induce ion-scale waves and generate shock surface ripples in supercritical shocks above the AIC critical Mach number, $M_{\rm AIC}^{*} \approx 2.3$.
Also electron-scale waves are generated primarily by the whistler instability in these high-$\beta$ shocks.
The resulting multi-scale waves from electron to ion scales are thought to be essential in the
electron injection to diffusive shock acceleration in $Q_{\perp}$-shocks in the ICM.

\end{abstract}


\section{Introduction} 
\label{sec:s1}

\begin{deluxetable*}{cccccccccc}[t]
\tablecaption{Linear Properties of the Instabilities driven by Perpendicular Temperature Anisotropies \label{tab:t1}}
\tabletypesize{\small}
\tablecolumns{4}
\tablenum{1}
\tablewidth{0pt}
\tablehead{
\colhead{instability} &
\colhead{AIC} &
\colhead{whistler} &
\colhead{ion-mirror} &
\colhead{electron-mirror} & 
}
\startdata
free energy source   & $T_{\rm i\perp}>T_{\rm i \parallel}$ & $T_{\rm e\perp}>T_{\rm e \parallel}$ & $T_{\rm i\perp}>T_{\rm i \parallel}$ & $T_{\rm e\perp}>T_{\rm e \parallel}$ \\
propagation angle with $\gamma_m$$^a$   & parallel & parallel & oblique & oblique \\
wavenumber& $ck/\omega_{\rm pi} \le 1$ & $ck/\omega_{\rm pe} \le 1$ & $ck/\omega_{\rm pi} \le 1$ & $ck/\omega_{\rm pe} \le 1$ \\
wave frequency& $0 <\omega_r < \Omega_{\rm ci}$ & $\Omega_{\rm ci} <\omega_r < \Omega_{\rm ce}$ & $\omega_r=0$ & $\omega_r=0$ \\
wave polarization    & LHCP$^b$ & RHCP$^b$ & Non-propagating  & Non-propagating 
\enddata
\tablenotetext{a}{$\gamma_m$ is the maximum growth rate.}
\tablenotetext{b}{LHCP (RHCP) stands for left-hand (right-hand) circular polarization.}
\end{deluxetable*}
 
Major mergers of galaxy clusters are known to drive weak shocks with sonic Mach numbers, $M_{\rm s} \lesssim 3$,
in the hot intracluster medium (ICM) of high $\beta$ \citep[e.g.,][]{ryu2003,skillman2008,vazza2009,hong2014,ha2018a}. 
Here, the plasma beta, $\beta = P_{\rm gas}/P_{\rm B}$, is the ratio of the gas pressure to the magnetic pressure.
The radiative signatures of such shocks have been detected in X-ray and radio observations \citep[e.g.,][]{bruggen2012,brunetti2014}.
In the case of so-called radio relics, the radio emission has been interpreted as the synchrotron radiation from  relativistic electrons accelerated via diffusive shock acceleration (DSA) in merger-driven shocks \citep[see][for a review]{vanweeren2019}.
 
To explain the origin of radio relics, this DSA model requires an electron preacceleration mechanism,
because postshock thermal electrons do not have momenta large enough to participate in the standard DSA process,
in which cosmic ray (CR) electrons diffuse across the shock \citep{kang2012}.
Since the width of the shock transition layer is comparable to the gyro-radius of postshock thermal ions, 
thermal electrons need to be energized to the so-called injection momentum, $p_{\rm inj} \sim {\rm a~few} \times p_{\rm i,th}$.
Here, $p_{\rm i, th} = \sqrt{2m_{\rm i}k_{\rm B}T_{\rm i2}}$ is the ion thermal momentum in the postshock gas of temperature $T_{\rm i2}$, $m_i$ is the ion mass, and $k_{\rm B}$ is the Boltzmann constant.
For shocks in the solar wind, the electron injection is observed preferentially at the 
quasi-perpendicular ($Q_{\perp}$) configuration with $\theta_{\rm Bn} \gtrsim 45^{\circ}$,
where $\theta_{\rm Bn}$ is the shock obliquity angle between the shock normal and the upstream
magnetic field direction \citep[e.g.,][]{gosling1989, oka2006, burgess2006}.

The electron preacceleration has been a key outstanding problem in understanding 
the production of CR electrons in weak ICM shocks.
Previous studies have shown that, in low-$M_{\rm s}$, high-$\beta$, $Q_{\perp}$-shocks, thermal electrons could be preaccelerated
primarily through the Fermi-like acceleration in the shock foot \citep{matsukiyo2011, guo2014a, guo2014b, kang2019} and the stochastic shock drift acceleration (SSDA)
in the shock transition region \citep{katou2019,niemiec2019,trotta2019}.
{\color{black}Although it has been shown that the electron preaccleration would be be enhanced by preexisting strong magnetic fluctuations in the low-$\beta$ ($\sim 1$) regime \citep[e.g.][]{guo2015,trotta2020}, the effect of such turbulence on high-$\beta$ ICM shocks has yet to be investigated and will not be considered here.

Both the Fermi-like acceleration and SSDA mechanisms} rely on the various microinstabilities triggered by the ion and electron temperature 
anisotropies in the shock structure \citep{gary1993}.
If $T_{\rm e\parallel}>T_{\rm e\perp}$, for example, the electron firehose instability (EFI) can grow with the following two branches:
the nonresonant, parallel-propagating mode with left-hand circular polarization, and the resonant, non-propagating, 
oblique mode \citep{gary2003}.
Hereafter, the subscripts $\parallel$ and $\perp$ denote the parallel and perpendicular directions to the background magnetic field, $\mathbf{B}_0$, respectively.
Under the condition of $T_{\rm e\perp}>T_{\rm e\parallel}$, by contrast, the whistler instability and
the electron-mirror (e-mirror) instability can be triggered \citep{scharer1967,gary1992,hellinger2018}.
The most unstable whistler mode propagates in the direction parallel to $\mathbf{B}_0$ with right-hand circular polarization,
while the e-mirror mode is non-propagating and has the maximum growth rate 
at the wavevector direction oblique with respect to $\mathbf{B}_0$.
In the case of $T_{\rm i\perp}>T_{\rm i\parallel}$, 
the Alfv\'{e}n ion cyclotron instability (AIC, or the proton cyclotron instability) and the ion-mirror (i-mirror) instability may become unstable \citep{winske1988, gary1993,gary1997,burgess2006}.
The fastest-growing mode of the AIC instability propagates in the direction parallel to $\mathbf{B}_0$ with left-hand circular polarization,
while the i-mirror mode is non-propagating and has the maximum growth rate at the wavevector direction oblique with 
respect to $\mathbf{B}_0$.
Table \ref{tab:t1} summarizes these linear properties of the instabilities driven by perpendicular temperature anisotropies, which are relevant for the present study.
 
In the foot of $Q_{\perp}$-shocks, the shock-reflected electrons backstream mainly along the upstream magnetic field
and induce an electron parallel anisotropy ($T_{\rm e\parallel}>T_{\rm e\perp}$), which could trigger the EFI and facilitate the Fermi-like preacceleration \citep{guo2014b, kang2019,kim2020}.
In the transition region behind the shock ramp, on the other hand, the AIC and i-mirror instabilities can be triggered by the ion perpendicular anisotropy ($T_{\rm i\perp}>T_{\rm i\parallel}$)
mainly due to the shock-reflected ions advected downstream, 
while the whistler and e-mirror instabilities can be excited by the electron perpendicular anisotropy 
($ T_{\rm e\perp}>T_{\rm e\parallel}$) mainly due to magnetic field compression at the shock ramp \citep{guo2017,katou2019}.
Such multi-scale waves from electron to ion scales are essential in the electron preacceleration via the SSDA \citep{matsukiyo2015, niemiec2019,trotta2019}.

Using two-dimensional (2D) particle-in-cell (PIC) simulations for $\beta\approx 20-100$, $Q_{\perp}$-shocks,
\citet{kang2019} showed that the Fermi-like preacceleration involving multiple cycles of shock drift acceleration (SDA) 
in the shock foot could be effective only in supercritical shocks with $M_{\rm s}$ greater than the EFI critical 
Mach number, $M_{\rm ef}^*\approx 2.3$.
However, they argued that the electron preacceleration may not proceed all the way to $p_{\rm inj}$,
because the EFI-driven waves are limited to electron scales.
\citet{niemiec2019}, on the other hand, performed a PIC simulation for $M_{\rm s}=3$ shock with $\beta=5$
in a 2D domain large enough to include ion-scale fluctuations, and suggested that electrons could be energized 
beyond $p_{\rm inj}$ via the SSDA
due to stochastic pitch-angle scattering off the multi-scale waves excited in the shock transition zone.

Furthermore, \citet{trotta2019} found that in $\beta\approx 1$ plasmas, the AIC instability is triggered 
and the ensuing electron preacceleration operates only in supercritical shocks
with the Alfv\'enic Mach number greater than the critical Mach number, $M_{\rm AIC}^*\approx 3.5$.
In a separate paper \citep[HKRK2021, hereafter]{ha2021}, 
we report a similar study of $\beta\approx 20-100$ shocks, which is design to explore through 2D PIC simulations 
how the multi-scale waves excited mainly by the AIC and whistler instabilities in the shock transition can assist 
the electron injection to DSA in ICM shocks.

In this paper, adopting the numerical estimates for the temperature anisotropies in the transition region
of the simulated shocks of HKRK2021\footnote{{\color{black} In HKRK2021 and hereafter, the transition zone is defined as the downstream region of $r_{\rm L,i}$ behind the shock ramp,
where $r_{\rm L,i}\approx u_0/\Omega_{\rm ci}^{\rm up}$ is the gyroradius of incoming ions; 
$u_0$ is the preshock flow speed defined in the downstream rest frame and $\Omega_{\rm ci}^{\rm up}$ is the gyro-frequency in the upstream.
Both the first and second overshoots and the accompanying undershoot are included in this zone, beyond which
the downstream states satisfy the canonical Rankine Hugoniot relation (see Figure 1 of HKRK2021).}},
we first perform a linear stability analysis for microinstabilities  
for wide ranges of parameters such as $M_{\rm s}=2-3$, $\beta=1-100$, and the ion-to-electron mass ratio,
$m_i/m_e=50-1836$.
{\color{black} This approach allows us to identify the most dominant modes of possible microinstabilities and 
to evaluate their linear properties for the set of realistic parameters pertaining to ICM shocks.
Hence, this kind of analyses on kinetic scales can provide crucial insights for
theoretical modelings and/or larger scale simulations of particle acceleration at weak high-$\beta$ shocks.
However, one of the limitations of PIC simulations is that it can follow kinetic plasma processes mainly at the low end of temporal and spatial scales 
owing to severe requirements of computational resources \citep[e.g.,][]{pohl2020}.}

In addition, adopting the same setup as in the linear analysis but only for the models with $\beta=50$ and $m_i/m_e=50$, 
we carry out 2D PIC simulations with periodic boundary conditions (periodic-box simulations, hereafter)
to study the nonlinear evolution of the plasma waves excited by such microinstabilities.
Note that throughout the paper we refer two different sets of PIC simulations:
(1) The `periodic-box simulations' are designed to study the nonlinear evolution of the excited plasma waves in the same set-up as in the linear analysis, 
and will be presented in Section \ref{sec:s3}.
(2) The `shock simulations' reported in HKRK2021 provide the numerical estimates for the ion and electron temperature anisotropies in the shock transition zone.

The paper is organized as follows. 
Section \ref{sec:s2} describes the linear analysis of the AIC, whistler, and mirror instabilities. 
In Section \ref{sec:s3}, we present the evolution of the waves driven by these instabilities in 2D periodic-box simulations.
In Section \ref{sec:s4}, the implication of our work on the shock criticality and shock surface ripples is discussed. 
A brief summary is given in Section \ref{sec:s5}.

\begin{figure}
\vskip -0.1 cm
\hskip -0.0 cm
\centerline{\includegraphics[width=0.5\textwidth]{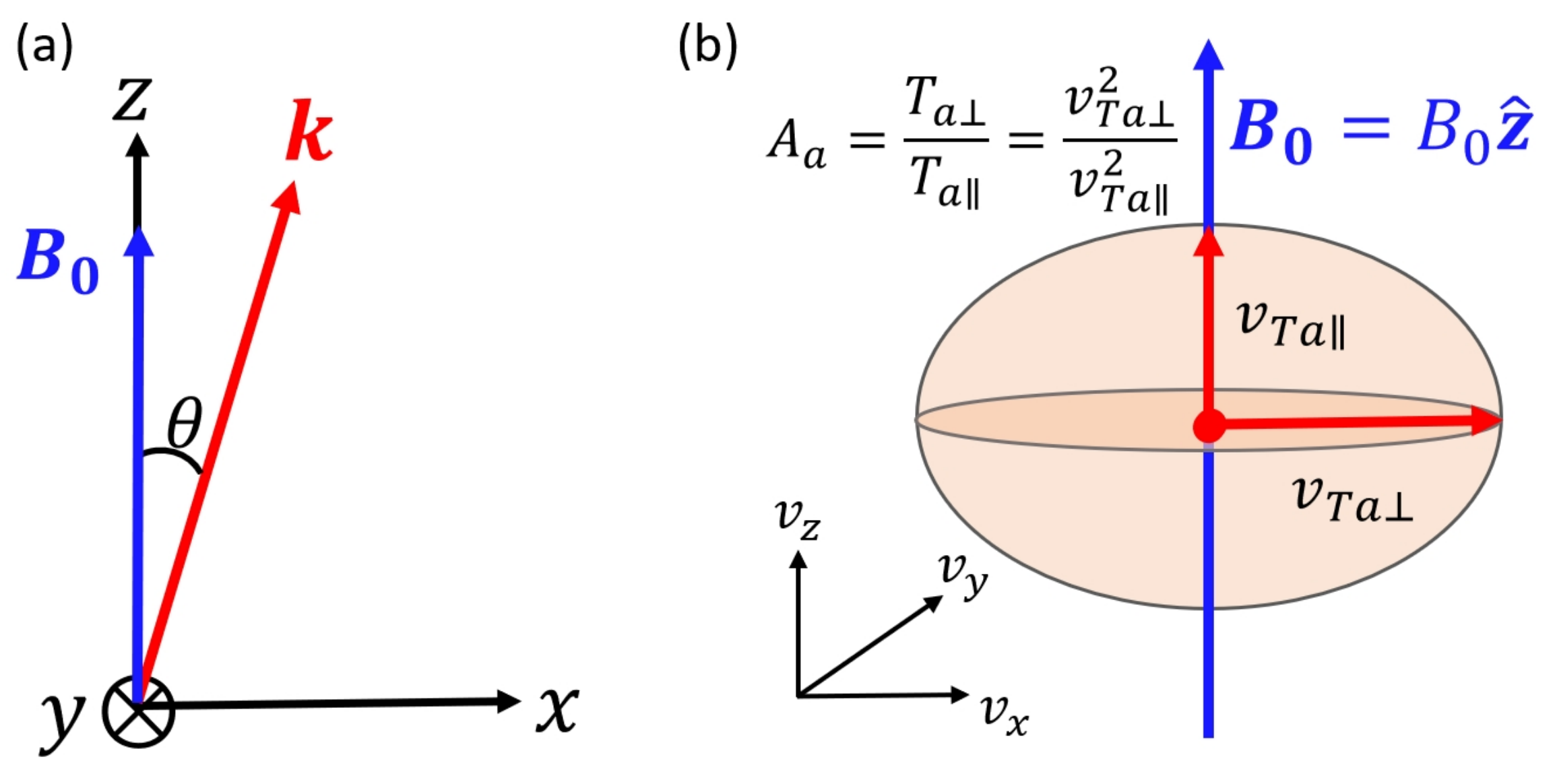}}
\vskip -0.1 cm
\caption{(a) Coordinate system employed in the present study. The background magnetic field, $\mathbf{B}_0 = B_0 \hat{z}$,
is parallel to the $+\hat{z}$ direction, while the wavevector, $\mathbf{k}=k_x \hat{x}+k_z \hat{z}$, lies in the
$x$-$z$ plane. The wave propagation angle, $\theta$, is the angle between $\mathbf{B}_0$ and $\mathbf{k}$.
(b) Schematic configuration showing the velocity ellipsoid of a bi-Maxwellian VDF with 
a temperature anisotropy $\mathcal{A}_a$, where $a$ denotes either `ion' or `electron'.
\label{fig:f1}}
\end{figure}

\section{Linear Analysis}
\label{sec:s2}

\subsection{Basic Equations}
\label{sec:s2.1}

We consider a homogeneous, collisionless, magnetized plasma, which is specified by the density and temperature of ions and electrons, $n_{\rm i}$, $n_{\rm e}$, $T_{\rm i}$, $T_{\rm e}$, and the background magnetic field $\mathbf{B}_0$. 
The linear dispersion relation of general electromagnetic (EM) modes is given as
\begin{equation}
\label{eq:e01}
\det \left(\epsilon_{ij}-\frac{c^2 k^2}{\omega^2}\big(\delta_{ij}-\frac{k_i k_j}{k^2}\big) \right)=0,
\end{equation}
where the dielectric tensor, $\epsilon_{ij}$, is determined by the plasma parameters and the velocity distribution functions (VDFs) of particles.  
Here, $k_i$ and $k_j$ are the components of the wavevector $\mathbf{k}$. 
Then, the complex frequency, $\omega = \omega_r + i \gamma$,\footnote{The quantity $i$ is the imaginary unit, not the coordinate component nor for ion species.} can be calculated as a function of the wavenumber, $k$, and the propagating angle, $\theta$, between $\mathbf{k}$ and $\mathbf{B}_0$. 
Without loss of generality, we set $\mathbf{B}_0 = B_0 \hat{z}$ along the $+z$ direction and $\bold{k}=k_x \hat{x} +k_z \hat{z}$ in the $x$-$z$ plane, as schematically illustrated in Figure \ref{fig:f1}(a). 

\begin{deluxetable*}{ccccccccccc}[t]
\tablecaption{Model Parameters and Linear Predictions \label{tab:t2}}
\tabletypesize{\small}
\tablecolumns{10}
\tablenum{2}
\tablewidth{0pt}
\tablehead{
\colhead{Model Name} &
\colhead{$M_{\rm s}$} &
\colhead{$\beta_{\rm e}$$^a$} &
\colhead{$\beta_{\rm i}$$^a$} &
\colhead{$\mathcal{A}_{\rm e}$$^a$} &
\colhead{$\mathcal{A}_{\rm i}$$^a$} &
\colhead{$m_{\rm i}/m_{\rm e}$} &
\colhead{AIC$^b$} &
\colhead{whistler$^c$} &
\colhead{ion-mirror$^b$} &
\colhead{electron-mirror$^c$} 
}
\startdata
LM2.0$\beta20$    & 2.0 & 9.7 & 10.3  & 1.1 & 1.1  & 50 & stable & stable & stable  & stable  \\
LM2.0$\beta50$     & 2.0 & 24 & 26  & 1.1 & 1.2  & 50 & stable  & (0.013,0.20,0) &  quasi-stable & (0.0029,0.14,69)  \\
LM2.0$\beta100$    & 2.0 & 48 & 52  & 1.1 & 1.2  & 50 & quasi-stable  & (0.035,0.20,0)  & (0.015,0.14,62)  & (0.008,0.15,56)  \\
\hline
LM2.3$\beta20$    & 2.3 & 8.4 & 12  & 1.1 & 1.5  & 50 & (0.041,0.21,0)  & (0.0016,0.26,0) & (0.04,0.28,63) & stable  \\
LM2.3$\beta50$       & 2.3 & 22 & 28  & 1.2 & 1.5  & 50 & (0.048,0.15,0)   & (0.03,0.24,0) & (0.054,0.21,61) & (0.0074,0.18,67)  \\
LM2.3$\beta100$    & 2.3 & 44 & 56  & 1.2 & 1.5  & 50 & (0.053,0.11,0)  & (0.056,0.24,0)  & (0.063,0.16,58) &  (0.016,0.17,61)   \\
\hline
LM3.0$\beta1$    & 3.0 & 0.48 & 0.52  & 1.2 & 1.2  & 50 & stable & stable & stable  & stable  \\
LM3.0$\beta5$    & 3.0 & 1.9 & 3.1  & 1.2 & 2.0  & 50 & (0.065,0.40,0) & (0.005,0.39,0) & (0.032,0.41,63) & stable  \\
LM3.0$\beta20$    & 3.0 & 7.5 & 13  & 1.2 & 2.0  & 50 &  (0.127,0.29,0)  &  (0.0156,0.30,0) &  (0.094,0.32,56)  &  (0.0016,0.17,74)  \\
LM3.0$\beta50$        & 3.0 & 19 & 31  & 1.2 & 2.0  & 50 & (0.145,0.20,0) & (0.059,0.29,0) & (0.11,0.23,55) & (0.015,0.22,64)  \\
LM3.0$\beta100$       & 3.0 & 38 & 62 & 1.2 & 2.0 & 50  &  (0.156,0.15,0) & (0.10,0.29,0) &  (0.12,0.18,54)  & (0.03,0.21,56) \\
\hline
LM3.0$\beta50$-m100       & 3.0 & 19 & 31  & 1.2 & 2.0  & 100 & (0.145,0.20,0) & (0.065,0.29,0) & (0.11,0.23,55) & (0.016,0.22,64) \\
LM3.0$\beta50$-m1836       & 3.0 & 19 & 31  & 1.2 & 2.0  & 1836 & (0.145,0.20,0) & (0.072,0.29,0) & (0.12,0.24,55) & (0.016,0.22,64) 
\enddata
\tablenotetext{a}{The quantities, $\beta_{\rm e}$, $\beta_{\rm i}$, $\mathcal{A}_{\rm e}$, and $\mathcal{A}_{\rm i}$, are obtained by the averaging numerical values over the transition zone in the simulated $Q_{\perp}$-shocks presented in HKRK2021.}
\tablenotetext{b}{Linear predictions for the fastest growing mode of the ion-driven instabilities, ($\gamma_m/\Omega_{\rm ci}$,$ck_m/\omega_{\rm pi}$,$\theta_m$), normalized with the ion gyro and plasma frequencies. $\theta_m$ is given in units of degree.}
\tablenotetext{c}{Linear predictions for the fastest growing mode of the electron-driven instabilities, ($\gamma_m/\Omega_{\rm ce}$,$ck_m/\omega_{\rm pe}$,$\theta_m$), normalized with the electron gyro and plasma frequencies. $\theta_m$ is given in units of degree.}
\end{deluxetable*}

In order to compute $\epsilon_{ij}$, we adopt the VDFs with bi-Maxwellian distributions for ions and electrons:
\begin{equation}
\label{eq:e02}
f_a(v_{\perp},v_{\parallel}) = \frac{n_0}{\pi^{3/2}v_{Ta \perp}^2 v_{Ta \parallel}} \exp\left(-\frac{v_{\perp}^2}{v_{Ta \perp}^2}-\frac{v_{\parallel}^2}{v_{Ta \parallel}^2}\right),
\end{equation}
where $v_{\perp} = \sqrt{v_x^2+v_y^2}$ and $v_{\parallel}=v_z$.
The subscript $a$ denotes $e$ or $i$ defined as the electron or ion species, respectively. 
Here, $n_0$ is the number density of electrons or ions, which satisfies the charge neutrality condition, i.e., $n_0=n_e=n_i$.
The parallel and perpendicular (to $\bold{B}_0$) thermal velocities are $v_{Ta \parallel}=\sqrt{2 k_{\rm B} T_{a \parallel}/m_a}$ and $v_{Ta \perp}=\sqrt{2 k_B T_{a \perp}/m_a}$, respectively. 
Then, the perpendicular temperature anisotropy of each particle species is given as $\mathcal{A}_a \equiv T_{a \perp}/T_{a \parallel}=v_{Ta \perp}^2/v_{Ta \parallel}^2$. 
The schematic configuration of  the thermal velocity ellipsoid for a bi-Maxwellian VDF with the temperature anisotropy $\mathcal{A}_a$ is shown in Figure \ref{fig:f1}(b).
As $\mathcal{A}_{a}$ increases, the thermal velocity surface in the velocity space deviates further away from the spherical shape.
Under these considerations, $\epsilon_{ij}$ is given as Equation (3) in \cite{kim2020} without the bulk drift velocities.
{\color{black}We note that in the shock simulations of HKRK2021, the VDFs of ions and electrons in the transition zone are likely non-gyrotropic and non-Maxwellian due to the SDA-reflected ions and electrons
accelerated via the gradient-$B$ drift (see Figure 4 of \citet{guo2014a}).
However, we expect the effects of non-Maxwellian VDFs on the linear predictions would be only marginal, 
since the fractions of particles in the suprathermal tail are order of $10^{-2}$ for ions and $\ll 10^{-2}$
for electrons in the downstream region of the fiducial $M_{\rm s}=3$ shock model (KRH2019).} 

For $n_0$, $\mathbf{B}_0$, $T_{a \parallel}$, and $T_{a \perp}$ of the homogeneous background plasma, 
we adopt the numerical values, averaged over the transition zone of
the simulated shocks of HKRK2021, where the preshock conditions are specified with the typical parameters 
of high-$\beta$ ICM plasmas, $n_{\rm ICM} = 10^{-4}{\rm cm^{-3}}$, $k_{\rm B}T_{\rm ICM} = (k_{\rm B}T_{\rm i} + k_{\rm B}T_{\rm e})/2 = 8.6$ KeV, and $\beta_{\rm ICM}=20-100$.
{\color{black}Again, in the shock simulations, both the ion and electron distributions are spatially 
nonuniform in the transition zone, where the flow structure oscillates with overshoots and undershoots 
in the longitudinal direction and ripples in the transverse direction.
Hence, we focus on qualitative analyses of the instabilities rather than making precise quantitative predictions.}

Throughout the paper, the plasma beta, $\beta_a=8\pi n_a k_B T_a/B_0^2$, the plasma frequency, $\omega_{\rm pa}^2=4\pi n_a e^2/m_a$, and the gyro-frequency, $\Omega_{\rm ca}=e B_0/m_a c$, for electrons and ions are used.
Note that in HKRK2021 the results are expressed in terms of the upstream parameters,
 $n_0^{\rm up}\approx n_0/r$ and $B_0^{\rm up}\approx B_0/r$, where $r$ is the density compression ratio across the shock ramp.
So for example $\omega_{\rm pa}^{\rm up}\approx \omega_{\rm pa}/\sqrt{r}$ 
and $\Omega_{\rm ca}^{\rm up}\approx \Omega_{\rm ca}/r$.

Plasma waves are characterized with the growth rate, $\gamma$, and the real frequency, $\omega_r$,
which are calculated by solving the dispersion relation in Equation (\ref{eq:e01}) for wavevector $\mathbf{k}$.
If the propagation angle of the wave with the maximum growth rate, $\gamma_m$, is $\theta_m\approx 0^{\circ}$,
the wave mode is called `parallel-propagating'. 
If $\theta_m \gg 0^{\circ}$, it is `oblique-propagating'.
If the wave frequency, $\omega_r\approx 0$, the mode is `non-propagating'.
The wave polarization, $P$, can be estimated also using the solution of the dispersion relation as follows:  
\begin{equation}
\label{eq:e03}
P \equiv {\rm sign} ({\omega_r}) \frac{|\delta E_+|-|\delta E_-|}{|\delta E_+|+|\delta E_-|},
\end{equation}
where $\delta E_{\pm} \equiv \delta E^{x}_{\mathbf{k},\omega} \mp i\delta E^{y}_{\mathbf{k},\omega}$ \citep{verscharen2013}.
The left-hand circular polarization (LHCP) corresponds to $P=-1$, whereas the right-hand circular polarization (RHCP) corresponds to $P=+1$.
Waves are in general elliptically polarized with $P \neq \pm 1$. 
In the case of non-propagating mode ($\omega_r=0$), $P=0$ (see Table \ref{tab:t1}).

\begin{figure*}[t]
\vskip -0.1 cm
\hskip -0.4 cm
\centerline{\includegraphics[width=0.8\textwidth]{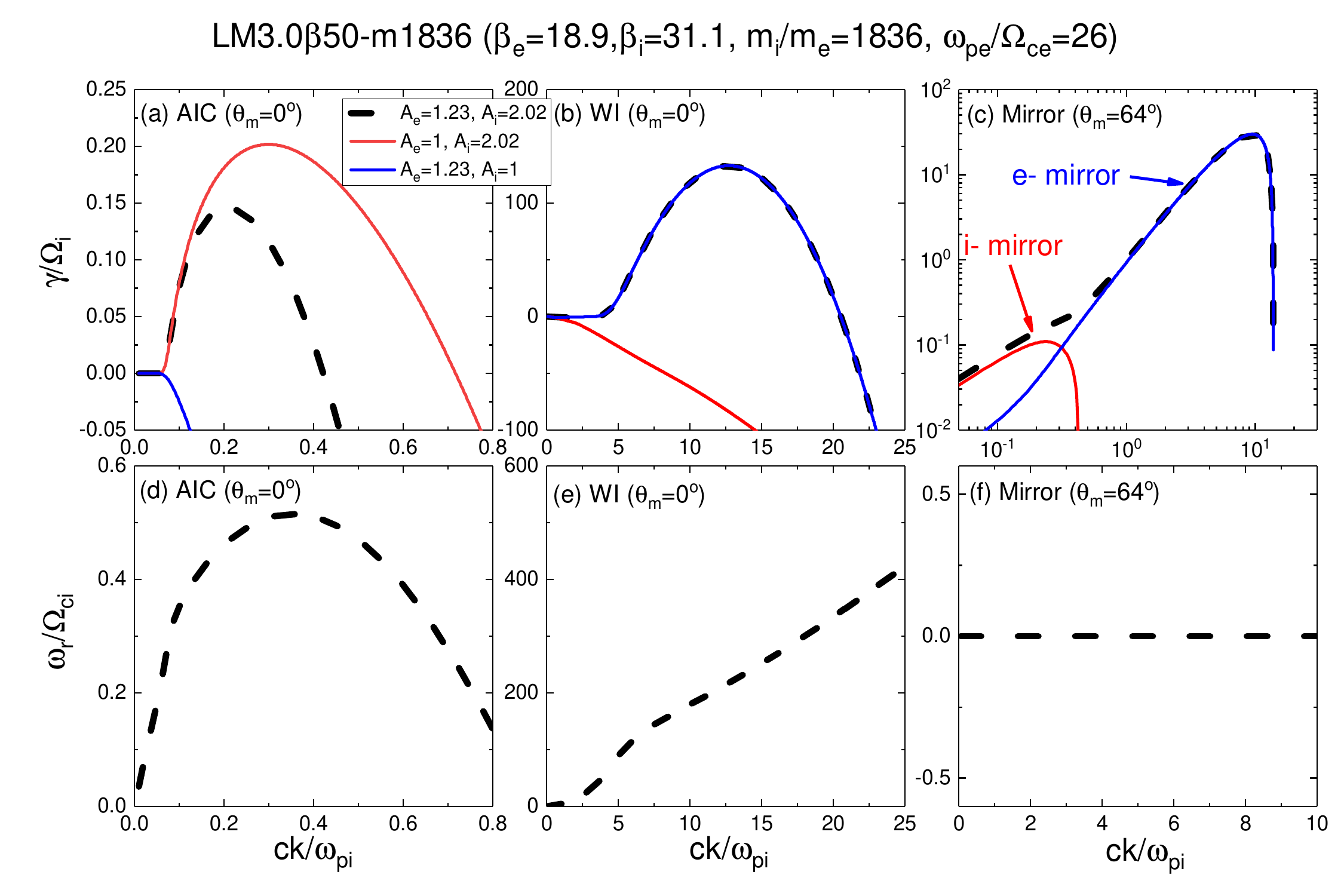}}
\vskip -0.3 cm
\caption{(a)-(c): Linear growth rate, $\gamma$, at the propagation angle of the fastest growing mode, $\theta_m$, 
for the AIC, whistler, and mirror modes, respectively, as a function of the wavenumber $k$ 
for the LM3.0$\beta50$-m1836 model. 
To examine separately the electron mode (blue) and the ion mode (red), the cases of $\mathcal{A}_{\rm e}=1.2$ and $\mathcal{A}_{\rm i}=1.0$ (blue) and $\mathcal{A}_{\rm e}=1.0$ and $\mathcal{A}_{\rm i}=2.0$ (red) are shown.
The black dashed lines show the mixed mode case, in which $\mathcal{A}_{\rm e}=1.2$ and $\mathcal{A}_{\rm i}=2.0$.
In panel (c) both $\gamma$ and $k$ are plotted in the logarithmic scales.
(d)-(f): Real frequency, $\omega_r$, 
for the same case as the black dashed lines in the upper panels.
Note that $\gamma$ and $\omega_r$ are normalized with $\Omega_{\rm ci}$ and 
$k$ is normalized with $\omega_{\rm pi}/c$, uniformly for both the ion and electron modes.
\label{fig:f2}}
\end{figure*}

\subsection{Linear Properties of AIC, Whistler and Mirror Instabilities}
\label{sec:s2.2}

In this section, we report the results of the linear stability analysis for the microinstabilities triggered by the ion and electron temperature anisotropies in the transition region of high-$\beta$, $Q_{\perp}$-shocks.
The first column of Table \ref{tab:t2} lists the model name, which is assigned with the two parameters,
the shock Mach number, $M_{\rm s}$, and $\beta$ ($\approx \beta_{\rm e}+\beta_{\rm i}$) in the shock transition region.
For example, the LM3.0$\beta$50 model has $M_{\rm s}=3.0$ and $\beta\approx 50$.
The values of the parameters, $\beta_{\rm e}$, $\beta_{\rm i}$, 
$\mathcal{A}_{\rm e}$, and $\mathcal{A}_{\rm i}$, are listed in the $3-6$ columns of the table.
For the models of $\beta\approx 20-100$ and the mass ration $m_i/m_e=50$, they are obtained with $n_0$, $\mathbf{B}_0$, $T_{a \parallel}$, and $T_{a \perp}$ estimated by averaging the numerical values over the transition region in
the simulated shocks with $M_{\rm s}=2-3$ and $\beta^{\rm up}= 20-100$ of HKRK2021.{\footnote{
Note that $\beta^{\rm up}$ for the shock models in HKRK2021 represents the plasma beta of the upstream, preshock plasmas,
while $\beta$ for the linear analysis models in Table \ref{tab:t2} is the plasma beta in the shock transition zone.
We found that $\beta\approx \beta^{\rm up}$ for the simulated shocks, although, in general, the plasma beta of the far downstream region is higher than $\beta^{\rm up}$.}}
Considering the uncertainties in averaging over nonlinear structures with overshoot/undershoot oscillations, 
they are given only up to two significant figures.

For the models with higher mass ratios, LM3.0$\beta50$-m100 with $m_i/m_e=100$ and LM3.0$\beta50$-m1836 with $m_i/m_e=1836$, 
the parameters for the LM3.0$\beta50$ model ($\beta_{\rm e}=19$, $\beta_{\rm i}=31$, 
$\mathcal{A}_{\rm e}=1.2$, and $\mathcal{A}_{\rm i}=2.0$) are used only for the linear analysis.
Also we carried out two additional shock simulations for M3.0$\beta1$ with $\beta=1$ and M3.0$\beta5$ with $\beta=5$, 
which were not considered in HKRK2021, in order to obtain the parameters to be used for 
LM3.0$\beta1$ and LM3.0$\beta5$.
Our fiducial models have $m_i/m_e=50$, which is adopted in order to ease
the requirements of computational resources for the periodic-box PIC simulations that will be described in Section \ref{sec:s3}. 

The linear predictions for the AIC, whistler, i-mirror, and e-mirror instabilities
are given in the $8-11$ columns of Table \ref{tab:t2}.
The three numbers inside each parenthesis present the linear properties of the fastest growing mode:  
($\gamma_m/\Omega_{\rm ci}$,$ck_m/\omega_{\rm pi}$,$\theta_m$) for the AIC and i-mirror instabilities, 
and ($\gamma_m/\Omega_{\rm ce}$,$ck_m/\omega_{\rm pe}$,$\theta_m$) for the whistler and e-mirror instabilities. 
Here, $k_m$ is the wavenumber that has the maximum growth rate $\gamma_m$ at $\theta_m$,
and $\theta_m$ is given in units of degree.
For a clear distinction between the ion and electron mirror modes, in the $10-11$ columns, $\gamma_m$ of each mirror mode,
obtained with either isotropic electrons (i.e., $\mathcal{A}_{\rm e}=1$, $\mathcal{A}_{\rm i}> 1$) or isotropic ions 
(i.e., $\mathcal{A}_{\rm i}=1$, $\mathcal{A}_{\rm e}> 1$), is shown.
Note that `stable' means that waves cannot grow because $\gamma_m <0$, and `quasi-stable' corresponds to $\gamma_m/\Omega_{\rm ci} <10^{-2}$.

Figure \ref{fig:f2} shows the linear analysis results for the LM3.0$\beta50$-m1836 model.
For the adopted parameters, $\omega_{\rm pe}/\Omega_{\rm ce}=26$.
Panels (a)-(c) display the growth (or damping) rate at $\theta_{m}$ of the AIC, whistler, and mirror instabilities, respectively, as a function of the wavenumber.  
To make a uniform comparison, $\gamma$ and $k$ are normalized with $\Omega_{\rm ci}$ and $\omega_{\rm pi}/c$, respectively,
for both the ion-driven and electron-driven instabilities.
Note that in panel (c) both $\gamma$ and $k$ are given in the logarithmic scales, in order to show
both the i-mirror and e-mirror modes in the same panel.
To examine the effects of $\mathcal{A}_{\rm i}$ and $\mathcal{A}_{\rm e}$ separately and also their combination, we present
the black dashed lines for the case with both the ion and electron anisotropies,
the red solid lines with the ion anisotropy only,
and the blue solid lines with the electron anisotropy only.

The AIC instability induces quasi-parallel modes with $\theta_m=0^{\circ}$.
Although $\mathcal{A}_{\rm i} > 1$ is the main free energy source which drives the AIC instability, 
we find that $\mathcal{A}_{\rm e} > 1$ reduces the growth rate (see the red and black lines in panel (a) and also  \citet{ahmadi2016}).
By contrast, the whistler instability is unstable for $\mathcal{A}_{\rm e} > 1$, and the growth rate is independent of $\mathcal{A}_{\rm i}$. 
The whistler mode is also quasi-parallel propagating with $\theta_m=0^{\circ}$.
The mirror modes, on the other hand, are highly oblique with $\theta_m=64^{\circ}$ for LM3.0$\beta50$-m1836. 
The e-mirror mode (blue) at high-$k$ ($ck/\omega_{\rm pi} > 0.3$) grows much faster than the i-mirror mode (red) at low-$k$ ($ck/\omega_{\rm pi} < 0.3$).
With both $\mathcal{A}_{\rm i} > 1$ and $\mathcal{A}_{\rm e} > 1$, a mixture of the two mirror modes appears in the intermediate-$k$ regime ($ck/\omega_{\rm pi} \sim 0.3$).

In the LM3.0$\beta50$-m1836 model, the maximum growth rates are given in the following order:
\begin{equation}
\label{eq:e04}
\gamma_{\rm WI} \gg \gamma_{\rm EM} \gg \gamma_{\rm AIC} > \gamma_{\rm IM},
\end{equation}
where $\gamma_{\rm WI}$, $\gamma_{\rm EM}$, $\gamma_{\rm AIC}$ and $\gamma_{\rm IM}$ are the maximum growth rates of the whistler, e-mirror, AIC and i-mirror instabilities, respectively.
Note that in general $\gamma_{\rm WI}>\gamma_{\rm EM}$ \citep{gary2006}, 
and $\gamma_{\rm AIC} > \gamma_{\rm IM}$ under space plasma conditions with low-$\beta$ and large temperature anisotropies \citep{gary1992,gary1993}.

The real frequency, $\omega_r/\Omega_{\rm ci}$, at $\theta_{m}$ for the mixed case ($\mathcal{A}_{\rm e}=1.2$ and $\mathcal{A}_{\rm i}=2.0$)
are shown in panels (d)-(f) of Figure \ref{fig:f2}. 
The AIC-driven mode has $\omega_r /\Omega_{\rm ci}\sim 0.25-0.5 $ for $ck/\omega_{\rm pi}\sim 0.1-0.4$,
while the whistler mode has $ \omega_r /\Omega_{\rm ci} \sim 80-350$ for $ck/\omega_{\rm pi}\sim 5-20$.
The mirror modes are non-propagating or purely growing with $\omega_r=0$.
Moreover, the polarization, calculated using the solutions of the dispersion relation, 
is $P= -1$, $+1$, and $0$ for the AIC, whistler, and
mirror instabilities, respectively, as expected. 

\begin{figure*}[t]
\vskip -0.1 cm
\hskip -0.4 cm
\centerline{\includegraphics[width=0.9\textwidth]{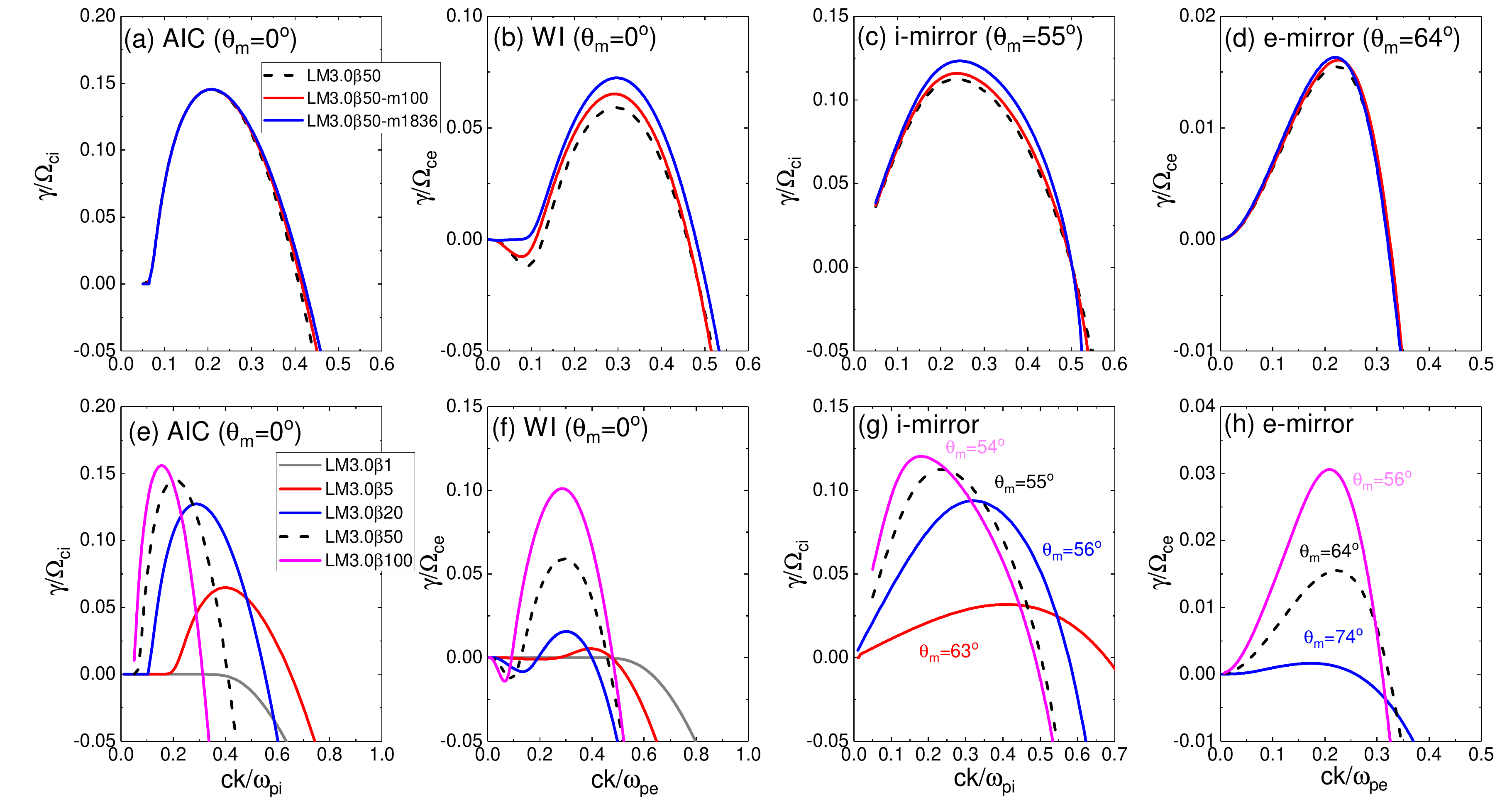}}
\vskip -0.15 cm
\caption{Dependence of the linear growth rate, $\gamma$, on $m_{\rm i}/m_{\rm e}$ (top) and $\beta$ (bottom); $\gamma$ at the propagation angle of the fastest growing mode, $\theta_m$, is given as a function of the wavenumber $k$.
The model parameters are listed in Table \ref{tab:t2}.
Note that for the mirror modes, $\theta_m$ depends on $\beta$, but not on $m_{\rm i}/m_{\rm e}$.
\label{fig:f3}}
\end{figure*}

\begin{figure*}[t]
\vskip -0.1 cm
\hskip -0.4 cm
\centerline{\includegraphics[width=0.9\textwidth]{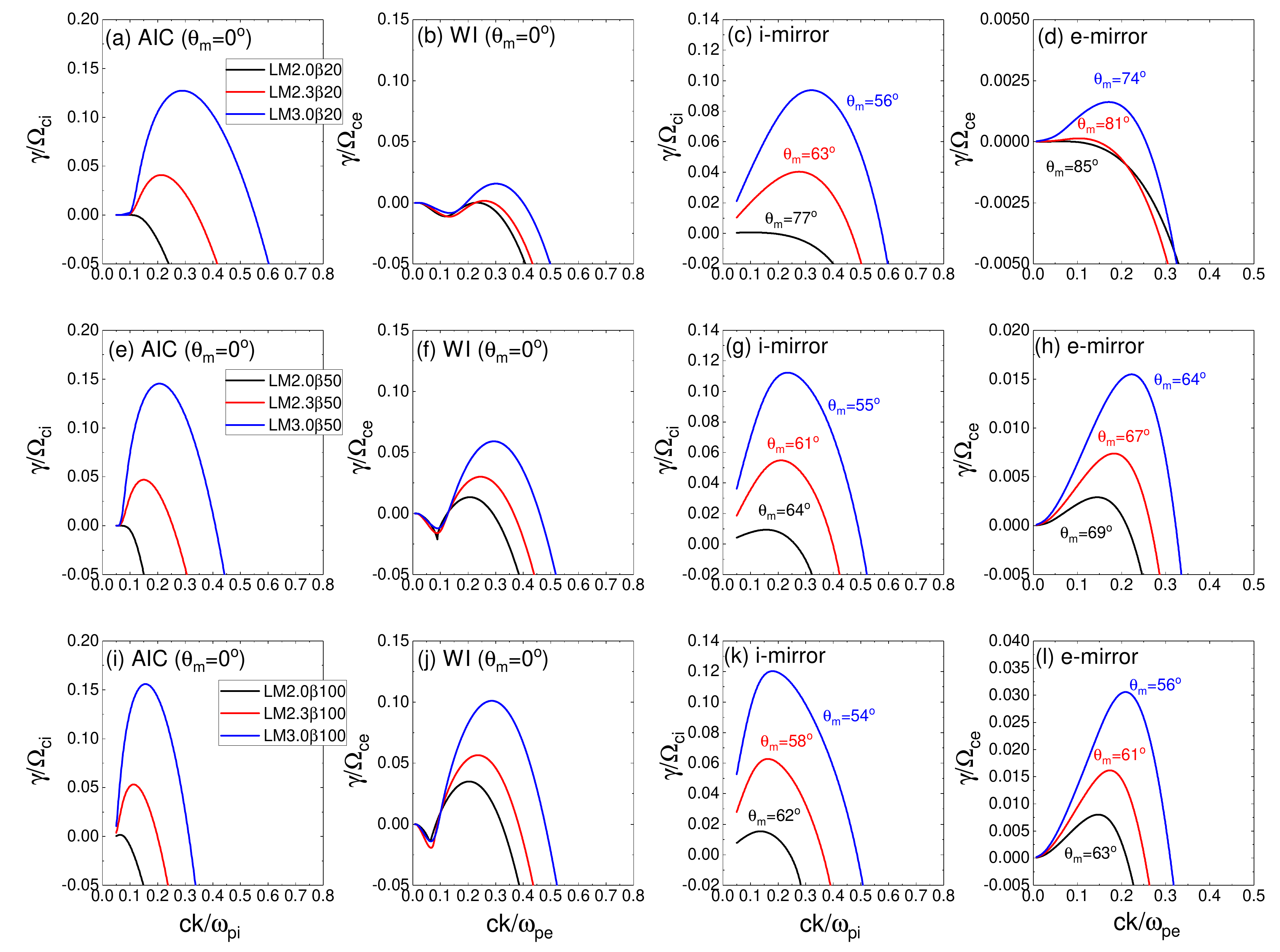}}
\vskip -0.25 cm
\caption{
Dependence of the linear growth rate, $\gamma$, on $M_{\rm s}$ and $\beta$; $\gamma$ at the propagation angle of the fastest growing mode, $\theta_m$, is given as a function of the wavenumber $k$.
In each panel, the black, red, and blue lines show the results for $M_{\rm s}=2.0,$ 2.3, and 3.0, respectively.
The plasma beta varies as $\beta=20$ (top), 50 (middle), and 100 (bottom).
The model parameters are listed in Table \ref{tab:t2}.
Note that for the mirror modes, $\theta_m$ depends on $\beta$.
\label{fig:f4}}
\end{figure*}

\subsection{Parameter Dependence of Linear Properties}
\label{sec:s2.3}

\begin{figure*}[t]
\vskip -0.1 cm
\hskip -0.4 cm
\centerline{\includegraphics[width=0.9\textwidth]{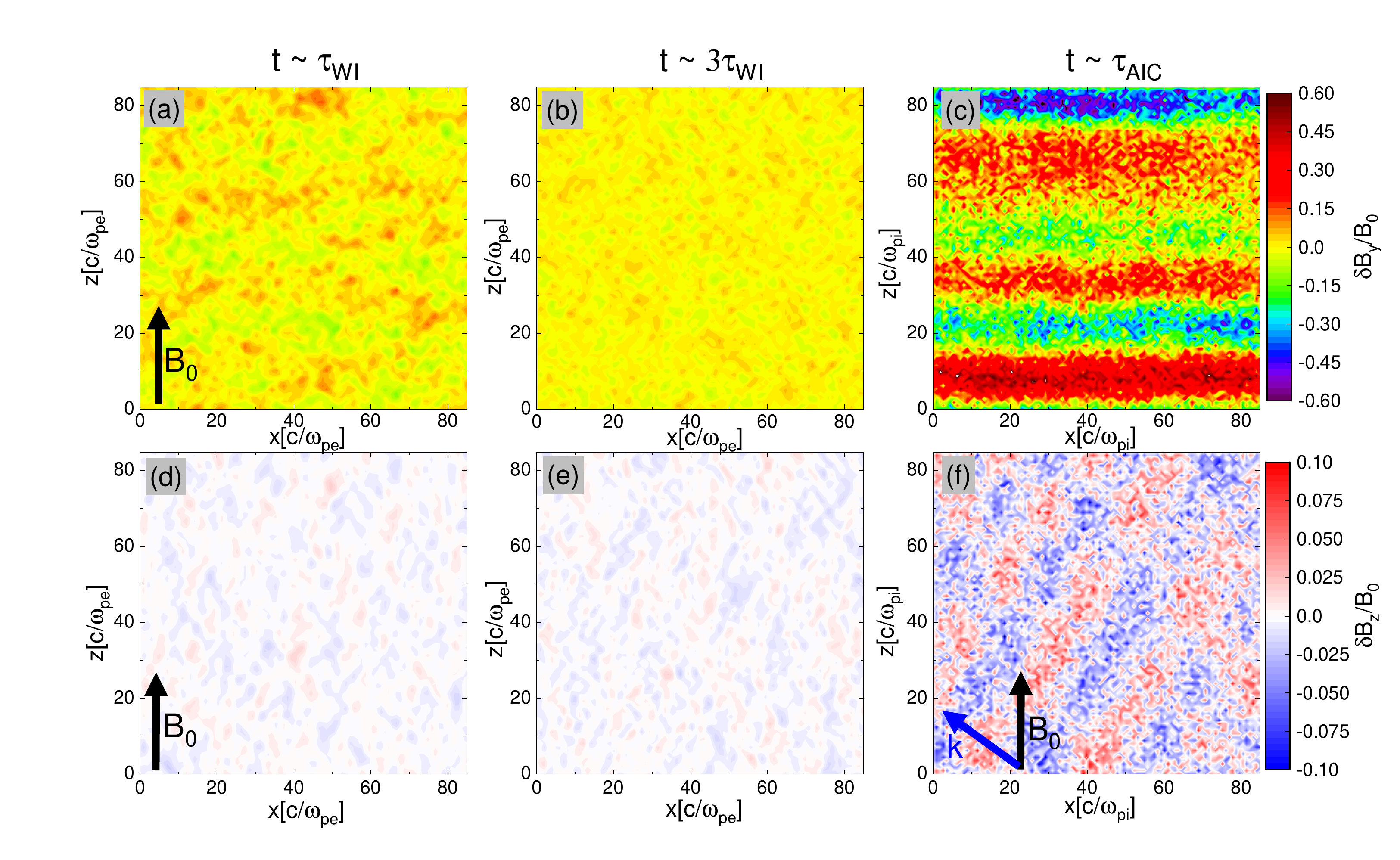}}
\vskip -0.4 cm
\caption{Magnetic field fluctuations, $\delta B_{y}$ (top) and $\delta B_{z}$ (bottom), 
in the periodic-box simulation for the LM3.0$\beta50$ model, plotted in the $x$-$z$ plane. 
At early times, $t\sim (1-3) \tau_{\rm WI}$, shown in panels (a)-(b) and (d)-(e), electron-scale waves are excited by the whistler and e-mirror instabilities, 
while ion-scale waves are generated by the AIC and i-mirror instabilities at $t\sim \tau_{\rm AIC}$ shown in panels (c) and (f). 
Note that the 2D domain with $[84.8\times 84.8] (c/\omega_{\rm pe})^2$ is shown in panels (a)-(b) and (d)-(e), while
the 2D domain with $[84.8\times 84.8] (c/\omega_{\rm pi})^2$ is shown in panels (c) and (f).
The black arrows indicate the direction of the background magnetic field, $\mathbf{B}_0$, 
while the blue arrow in panel (f) shows the direction of wave propagation, $\mathbf{k}$, for the i-mirror mode 
with the maximum growth rate.  
\label{fig:f5}}
\end{figure*}

As listed in Table \ref{tab:t2}, we consider a number of models 
to explore the dependence on $m_{\rm i}/m_{\rm e}$ and $\beta$.
The upper panels of Figure \ref{fig:f3} show the linear predictions for the models with
 $M_{\rm s}=3$, $\beta=50$, and $m_{\rm i}/m_{\rm e}=50-1836$,
while the lower panels are for the models with $M_{\rm s}=3$, $m_i/m_e=50$, and $\beta=1-100$.
For a higher mass ratio, electrons go through more gyro-motions per the ion gyro-time, $\Omega_{\rm ci}^{-1}$. 
Nevertheless, $\gamma_{\rm AIC}/\Omega_{\rm ci}$ and $\gamma_{\rm EM}/\Omega_{\rm ce}$ are almost independent of $m_{\rm i}/m_{\rm e}$.
In the case of the whistler and i-mirror instabilities, on the other hand, overall, the normalized growth rates are slightly lower for smaller $m_i/m_e$.
Also the damping rate for the whistler instability is slightly higher for smaller $m_i/m_e$ in the small wavenumber regime ($ck/\omega_{\rm pe} \sim 0.1$). 
As a result, the growth of the whistler and i-mirror instabilities may be somewhat suppressed
in the shock simulations with reduced mass ratios.
However, even in the case of $m_{\rm i}/m_{\rm e}=50$, this effect is expected to be only minor, 
because the inequality in Equation (\ref{eq:e04}) is still valid and the changes of $k_m$ and $\theta_m$ are negligible (see Table \ref{tab:t2}).

The plasma beta is another important parameter that affects the stability of the system.
Note that the anisotropy parameters, $\mathcal{A}_{\rm e}$ and $\mathcal{A}_{\rm i}$, are almost independent of $\beta$ for $\beta\approx 20-100$, the range relevant for ICM shocks (see Table \ref{tab:t2}), although they tend to increase slightly
with increasing $\beta$ in the second digit to the right of the decimal point.
In the low-$\beta$ case (LM3.0$\beta$1), $\mathcal{A}_{\rm i}=1.2$ is significantly smaller than those of other high-$\beta$ models due to the strong magnetization of ions.
This is because $\mathcal{A}_{\rm i}$ in the shock transition is closely related to the fraction of reflected ions.
On the other hand, $\mathcal{A}_{\rm e}$ in the shock transition is not substantially affected by $\beta$, because it is mainly determined by the magnetic field compression rather than the fraction of reflected electrons.
Given the same temperature anisotropies,
the growth of the instabilities tends to be suppressed by strong magnetic fields at low-$\beta$ plasmas.  
As can be seen in the lower panels of Figure \ref{fig:f3},
the peak values of either $\gamma_m/\Omega_{\rm ci}$ for the ion-driven modes or $\gamma_m/\Omega_{\rm ce}$ for the electron-driven modes
increase with increasing $\beta$.
For the AIC, whistler, and i-mirror instabilities, $\gamma_m/\Omega_{\rm ci}$ or $\gamma_m/\Omega_{\rm ce}$ occurs at smaller $ck/\omega_{\rm pi}$ or $ck/\omega_{\rm pe}$, for higher $\beta$. 
But such a trend is not obvious in the case of the e-mirror mode.

In the high-$\beta$ cases ($\beta\approx 20-100$) with $M_{\rm s}=3$, all the AIC, whistler, i-mirror and e-mirror waves can be triggered, as shown in the lower panels of Figure \ref{fig:f3}, leading to
the generation of multi-scale waves from electron to ion scales.
On the other hand, in the LM3.0$\beta$5 model (red solid lines), the e-mirror mode is stable, but other modes are unstable. 
In the LM3.0$\beta$1 model (gray solid lines), all the instabilities are stable with negative growth rates.

The sonic Mach number, $M_{\rm s}$, is the key parameter that determines the temperature anisotropies in the transition of 
high-$\beta$ ICM shocks ($\beta\approx 20-100$), 
since the ion reflection fraction and the magnetic field compression are closely related to $M_{\rm s}$.
Figure \ref{fig:f4} shows the growth rates of the instabilities for $M_{\rm s}=2.0$ (black), 2.3 (red), and 3.0 (blue),
in the cases of $\beta=20$ (top), 50 (middle) and 100 (bottom).
As $M_{\rm s}$ increases, both $\mathcal{A}_{\rm e}$ and $\mathcal{A}_{\rm i}$ increase, so all the modes grow faster and 
$k_m$ shifts towards larger $k$, regardless of $\beta$.

Note that the AIC and whistler modes have $\gamma_m$ at $\theta_m=0$ independent of $M_{\rm s}$, 
whereas $\theta_m$ decreases with increasing $M_{\rm s}$ for the i-mirror and e-mirror modes (see also Table \ref{tab:t2}).
In LM2.0$\beta$50 and LM2.0$\beta$100, the AIC instability is stable or quasi-stable,
while the whistler and mirror modes can grow.
In the case of LM2.0$\beta$20, all the instabilities are stable (see black lines in top panels).  
In the models with $M_{\rm s}=2.3-3$ (red and blue lines), on the other hand, the four instabilities are unstable, and hence
multi-scale plasma waves can be generated.

{\color{black}
The parameter dependence can be summarized as follows.
(1) For the AIC mode, the maximum {\it normalized} growth rate, $\gamma_m/\Omega_{\rm ca}$, and the corresponding {\it normalized} wavenumber, $ck_m /\omega_{\rm pa}$, are almost independent of $m_i/m_e$. 
For the whistler, i-mirror, and e-mirror modes, on the other hand, $\gamma_m/\Omega_{\rm ca}$ is only slightly enhanced for larger $m_i/m_e$, whereas $ck_m /\omega_{\rm pa}$ exhibits almost no dependence.
(2) For all the modes, the overall trend shows that $\gamma_m/\Omega_{\rm ca}$ is higher and $ck_m /\omega_{\rm pa}$ is smaller for higher $\beta$.
(3) For all the modes, $\gamma_m/\Omega_{\rm ca}$ is higher and $ck_m /\omega_{\rm pa}$ is larger for higher $M_{\rm s}$ cases with larger $\mathcal{A}_{\rm e}$ and larger $\mathcal{A}_{\rm i}$.
}

\section{Nonlinear Evolution of Induced Waves in Periodic-Box Simulations}
\label{sec:s3}

\subsection{Numerical Setup}
\label{sec:s3.1}

To investigate the development and nonlinear evolution of the instabilities, we performed 2D PIC simulations with periodic boundary conditions
for the three fiducial models, LM2.0$\beta50$, LM2.3$\beta50$ and LM3.0$\beta50$,
with the same setup described in Section \ref{sec:s2.1}.
Electrons and ions are prescribed with bi-Maxwellian VDFs with $\beta_{\rm e}$, $\beta_{\rm i}$, $\mathcal{A}_{\rm e}$,
$\mathcal{A}_{\rm i}$ given in Table \ref{tab:t2}.
As noted before, here $m_{\rm i}/m_{\rm e} = 50$ is employed due to the computational limitations, but at least the early, linear-stage development of the plasma instabilities under consideration is expected to depend rather weakly on the mass ratio.

{\color{black}
We point that the setup for these 2D periodic-box simulations should intrinsically differ from the condition in the transition zone of shocks in the following aspects.
(1) The initial distributions of ions and electrons are allowed to relax in the periodic-box simulations.
By contrast, the shock-reflected ions and electrons are continuously convected into the transition zone behind the shock ramp, leading to the continuous excitation of the instabilities.
(2) Homogeneous spatial distributions and bi-Maxwellian VDFs are assumed for the periodic-box simulations. 
On the other hand, as noted in Section \ref{sec:s2.1}, both the ion and electron distributions are spatially nonuniform 
and the VDFs of ions and electrons contain suprathermal tails in the shock transition region. 
Nevertheless, the periodic-box simulations like ours are often used to investigate the nonlinear evolution and properties of microinstabilities in the either upstream or downstream region near the shock \citep[e.g.][]{scholer2000,guo2014b,trotta2020}.}

\begin{figure*}[t]
\vskip -0.1 cm
\hskip -0.5 cm
\centerline{\includegraphics[width=0.85\textwidth]{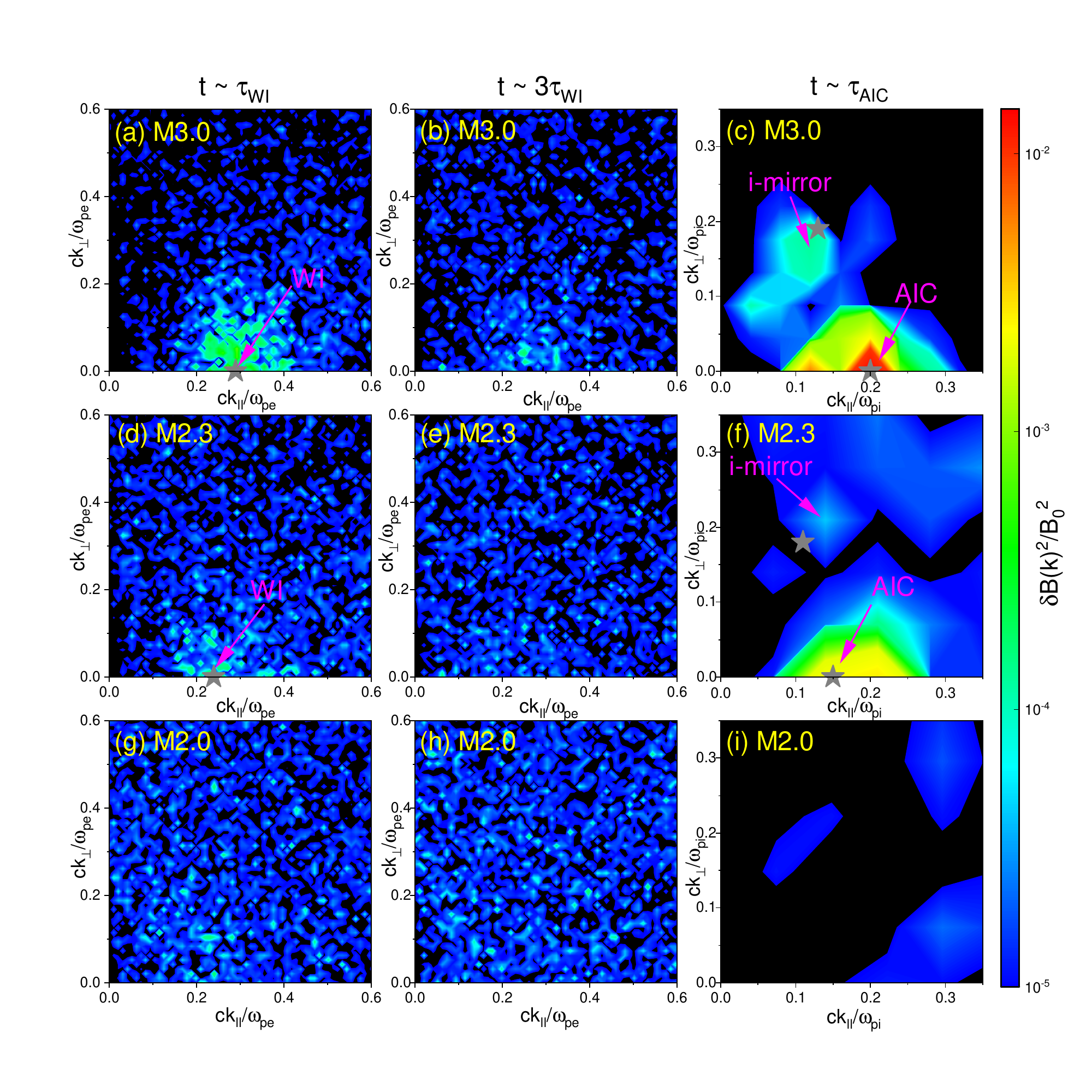}}
\vskip -0.9 cm
\caption{Power spectra of the magnetic field fluctuations, $\delta B_{y}^{2} (\bold{k})$, in the period-box simulations
for LM3.0$\beta$50 (top), LM2.3$\beta$50 (middle), and LM2.0$\beta$50 (bottom), plotted in the $k_{\parallel}$-$k_{\perp}$ (that is, $k_z-k_x$) plane. 
The results are shown at $t\sim \tau_{\rm WI}$ (left), $t\sim 3 \tau_{\rm WI}$ (middle), and $t\sim \tau_{\rm AIC}$ (right).
See the text for the remarks on $\tau_{\rm AIC}$ for LM2.0$\beta$50.
The gray star symbol marks the location of the maximum linear growth rate, $\gamma_m$, estimated from the linear analysis.
In the models with $M_{\rm s} \geq 2.3$, AIC, whistler and i-mirror waves appear, while those waves do not grow substantially in the model with $M_{\rm s}= 2$. 
\label{fig:f6}}
\end{figure*}
  
The simulations were carried out using a parallelized EM PIC code, TRISTAN-MP \citep[]{buneman1993,spitkovsky2005}.
The simulation domain is a square of box size $L_{\rm x} = L_{\rm z} = 84.8 c/\omega_{\rm pi}= 600 c/\omega_{\rm pe}$ in the 
$x$-$z$ plane, which consists of the grid cells of $\Delta x = \Delta z = 0.1 c/\omega_{\rm pe}$. 
In each cell, 32 particles (16 for ions and 16 for electrons) are placed.
The time step of the simulations is $\Delta t = 0.045 \omega_{\rm pe}^{-1}$, and the simulations ran up to $t_{\rm end} = 130 \Omega_{\rm ci}^{-1}$. 
{\color{black}Interpreting the results of our PIC simulations could be limited by numerical noises and aliases 
due to a finite number of macroparticles on discretized grids \citep[e.g.][]{pohl2020}.
However, we expect that the overall results of the PIC simulations are reasonably converged, judging from the previous
studies \citep{guo2014b,kang2019,kim2020}.}

\subsection{Results of Periodic-Box Simulations}
\label{sec:s3.2}

With the inequality in Equation (\ref{eq:e04}), we expect that
the whistler mode grows much faster than other modes,
resulting in the relaxation of $\mathcal{A}_{\rm e}$ during the early stage.
As the whistler and e-mirror modes grow and then decay on the time scale of $\tau_{\rm WI} \equiv 1/\gamma_{\rm WI}$, 
the AIC and i-mirror modes become dominant later on the time scale of $\tau_{\rm AIC} \equiv 1/\gamma_{\rm AIC}$.

Figure \ref{fig:f5} shows the magnetic field fluctuations, $\delta B_{y}$ (upper panels) and $\delta B_{z}$ (lower panels),
in the $x$-$z$ plane (simulation plane) at three different times in the LM3.0$\beta50$ model.
Here, the growth time scales, $\tau_{\rm WI}$ and $\tau_{\rm AIC}$, are estimated with $\gamma_m$ of each mode in Table \ref{tab:t2}.
At $t \sim \tau_{\rm WI}$, the transverse component, $\delta B_{y}$, appears on electron scales and the waves containing it propagate parallel to $\mathbf{B}_{0}$ in panel (a), but the longitudinal component, $\delta B_{z}$, does not grow significantly in panel (d).
In this early stage, the dominant mode is the whistler mode,
while the e-mirror mode is much weak to be clearly manifested.
As $\mathcal{A}_{\rm e}$ decreases in time due to the electron scattering off the excited waves, the whistler waves decay as shown in panel (b). 
On the time scale of $\tau_{\rm AIC}$, both the AIC and i-mirror instabilities grow and become dominant.  
It is clear that the AIC-driven waves, shown in panel (c), are parallel-propagating,
while the i-mirror-driven waves, shown in panel (f), are oblique-propagating;
the blue arrow in the bottom-left corner of panel (f) denotes the wavevector of the i-mirror-driven mode with the maximum growth rate.

\begin{deluxetable*}{ccccccccccc}[t]
\tablecaption{Shock Criticality of the Simulated Shock Models and Stability of the Linear Analysis Models \label{tab:t3}}
\tabletypesize{\small}
\tablecolumns{10}
\tablenum{3}
\tablewidth{0pt}
\tablehead{
\colhead{Simulated Shock Model} &
\colhead{Shock Criticality} &
\colhead{Linear Analysis Model} &
\colhead{AIC} &
\colhead{WI} &
\colhead{ion-mirror} &
\colhead{electron-mirror} 
}
\startdata
M2.0$\beta20$    & sub & LM2.0$\beta20$  & stable & stable & stable  & stable  \\
M2.0$\beta50$     & sub & LM2.0$\beta50$ & stable & unstable & quasi-stable & unstable  \\
M2.0$\beta100$    & sub & LM2.0$\beta100$ & quasi-stable & unstable  & unstable  & unstable  \\
\hline
M2.3$\beta20$    & super & LM2.3$\beta20$  & unstable & unstable & unstable & stable \\
M2.3$\beta50$      & super & LM2.3$\beta50$ & unstable & unstable & unstable & unstable \\
M2.3$\beta100$    & super & LM2.3$\beta100$ & unstable & unstable & unstable & unstable \\
\hline
M3.0$\beta1$    & sub & LM3.0$\beta1$ & stable & stable & stable  & stable  \\
M3.0$\beta5$    & super & LM3.0$\beta5$ & unstable & unstable & unstable & stable  \\
M3.0$\beta20$    & super & LM3.0$\beta20$ &  unstable & unstable &  unstable  & unstable  \\
M3.0$\beta50$      & super & LM3.0$\beta50$ & unstable & unstable & unstable & unstable \\
M3.0$\beta100$    & super & LM3.0$\beta100$ & unstable & unstable & unstable & unstable
\enddata
\end{deluxetable*}

Figure \ref{fig:f6} shows the time evolution of the power spectrum for the magnetic field fluctuations, $\delta B_{y}^{2} (\bold{k})$, for LM2.0$\beta50$, LM2.3$\beta50$, and LM3.0$\beta50$ at $t\sim \tau_{\rm WI}$, $t\sim 3 \tau_{\rm WI}$, and $t\sim \tau_{\rm AIC}$ .
Again, the growth time scale of each mode is estimated with $\gamma_m$ listed in Table \ref{tab:t2},
except for the LM2.0$\beta50$ model, in which the AIC instability is stable, 
and so the output time of panel (i) is chosen at the evolutionary stage similar to that of LM2.3$\beta50$.
In the cases of $M_{\rm s}=2.3$ and $3$, whistler waves are excited dominantly at quasi-parallel propagating angles at $t\sim \tau_{\rm WI}$.
After the initial linear stage, the energy of the whistler waves is transferred to smaller wavenumbers and the waves gradually decay, as shown in panels (b) and (e).
On the time scale of $\sim\tau_{\rm AIC}$, AIC waves and i-mirror waves appear dominantly 
at quasi-parallel and highly oblique angles, respectively, as shown in panels (c) and (f). 
This is consistent with the evolutionary behavior which we have described with Figure \ref{fig:f5}.
For the AIC and whistler instabilities, the linear predictions for $\mathbf{k_m}$ with the maximum growth rate (gray star symbols)
agree reasonably well with the peak locations of the magnetic power spectrum realized in the PIC simulations.
But the linear estimates for the i-mirror mode are slightly off, because $\gamma_m$ is obtained 
without the electron anisotropy, as stated in Section \ref{sec:s2.2}.
In summary, the results of the periodic-box simulations are quite consistent with the linear predictions described earlier.
Also we note that the results of our PIC simulations are in good agreement with those of \cite{ahmadi2016},
in which PIC simulations were carried out to explore the evolution of the instabilities due to the temperature anisotropies
in space plasmas with $\beta \sim 1$.
The bottom panels of Figure \ref{fig:f6} confirm that waves do not grow noticeably in the LM2.0$\beta$50 model.

In these periodic-box simulations, 
the electron-scale waves develop first and then decay as $\mathcal{A}_{\rm e}$ is relaxed in the early stage, followed by
the growth of the ion-scale waves due to $\mathcal{A}_{\rm i}$.
In the shock transition region, by contrast, temperature anisotropies are to be supplied continuously by newly reflected-gyrating ions and magnetic field compression,   
hence multi-scale plasma waves from electron to ion scales are expected to be simultaneously present.

\section{Implications for Shock Simulations}
\label{sec:s4}

\subsection{Shock Criticality}
\label{sec:s4.1}
As mentioned in the introduction,
the Fermi-like acceleration, which relies on the upstream waves excited by the EFI, is effective only for supercritical shocks
with $M_{\rm s}\ge M_{\rm EFI}^* \approx 2.3$ in $\beta\approx 20-100$ plasma \citep{guo2014b,kang2019}.
The SSDA, which depends on the multi-scale waves excited mainly by the AIC and whistler instabilities, is thought to occur in supercritical shocks 
with $M_{\rm s}\ge M_{\rm AIC}^* \approx 3.5$ in $\beta\approx 1$ plasmas \citep{trotta2019}
and $M_{\rm s}\ge M_{\rm AIC}^* \approx 2.3$ in $\beta\approx 20-100$ plasmas (HKRK2021).
We suggest that both $M_{\rm EFI}^*$ and $M_{\rm AIC}^*$ are related to the sonic critical Mach number, $M_{\rm s}^*$, for ion reflection,
since the structure of collisionless shocks is governed primarily by the dynamics of shock-reflected ions.

Table \ref{tab:t3} summarizes
the shock criticality of the simulated shock models 
and the stability of the linear analysis models.
The first column lists the name of the simulated shock models considered in HKRK2021, and the two additional models for low-$\beta$ shocks performed for this study.
The shock criticality of each model is given in the second column.
The name of the corresponding linear analysis models is given in the third column, while the last four columns
show the stability for the four instabilities (see also Table \ref{tab:t2}).
We note that the name of the shock models includes $\beta^{\rm up}$ in the preshock, upstream plasmas,
while that of the linear analysis models includes $\beta$ in the shock transition zone given in Table \ref{tab:t2}.

As discussed in Sections \ref{sec:s2} and \ref{sec:s3}, 
in $\beta\approx 20-100$ plasmas, the AIC instability operates for $M_{\rm s} \gtrsim 2.3$,
while whistler waves are induced regardless of $M_{\rm s}$.  
In the M3.0$\beta$5 model, the e-mirror mode is stable, while the other three modes are unstable.
This is in good agreement with the 2D simulation of a $M_s=5$ and $\beta=5$ shock reported earlier by \citet{niemiec2019}.
In the M3.0$\beta$1 model, by contrast, $\mathcal{A}_{\rm i}$ is smaller than that of high-$\beta$ models,
and all the four instability modes are suppressed by strong magnetization. 
This is consistent with the results of $M_{\rm AIC}^*\approx 3.5$ for shocks with $\beta\approx 1$ presented by \citet{trotta2019}.

\subsection{Shock Surface Rippling}
\label{sec:s4.1}

Another important feature of supercritical shocks above $M_{\rm AIC}^*$ is the shock surface rippling. 
According to previous shock simulations \citep[e.g.,][]{lowe2003,matsukiyo2015, niemiec2019, trotta2019}, 
the rippling has the characters of AIC waves with the fastest growing mode at $\theta_m\sim 0$, the propagation
speed close to the local Alfv\'en speed, and the wavelengths of $\sim \lambda_{\rm AIC}$ ($\approx 30c/\omega_{\rm pi}$).
{\color{black}In fact, shock ripples have been observed at the Earth's bow shock 
and the interplanetary shocks inside the heliosphere, and investigated extensively in space physics \citep[e.g.][]{winske1988, moullard2006,johlander2016}. 
At the interplanetary shocks, ripples on scales even larger than $\lambda_{\rm AIC}$ have been detected as well,
and are thought to be triggered by the upstream magnetic structures produced by backstreaming ions \citep{kajdic2019}.} 

The parallel-propagating AIC and whistler waves in homogeneous plasmas are purely electromagnetic and incompressible with both the electric and magnetic wave vectors pointing normal to $\mathbf{B}_{0}$.
The fluctuating magnetic fields of oblique mirror modes, on the other hand, have a substantial 
longitudinal component, that is, $\delta \mathbf{B}$ has a significant component parallel to $\mathbf{B}_0$ \citep{gary1993}. 
Since the density fluctuations are proportional to the parallel electric and magnetic field fluctuations \citep{hojo1993},
we expect to see only weak ion density fluctuations due to the i-mirror mode in our 2D periodic-box simulations.

\begin{figure}[t]
\vskip -0.5 cm
\hskip -0.0 cm
\centerline{\includegraphics[width=0.45\textwidth]{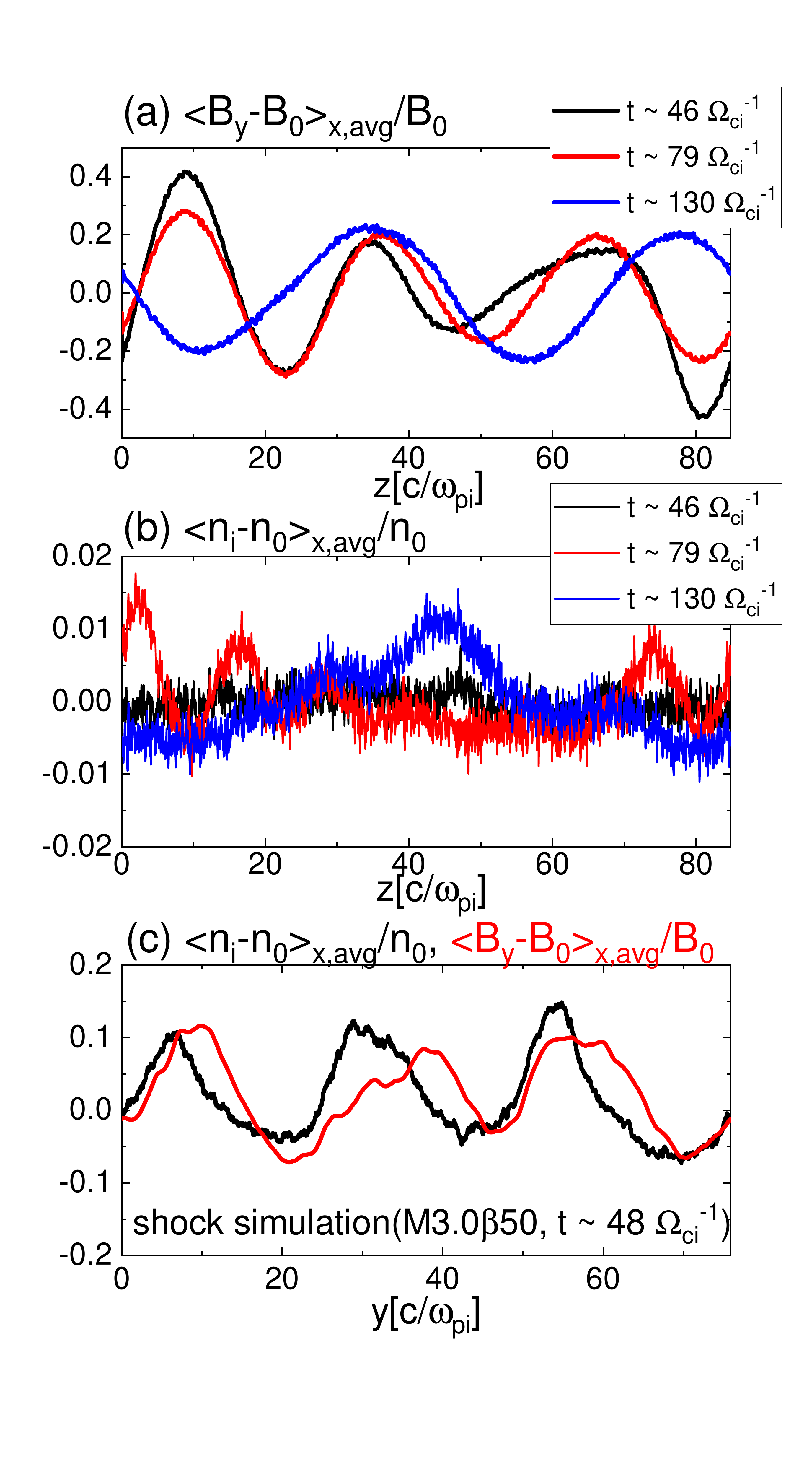}}
\vskip -1.5 cm
\caption{(a)-(b): Variations in the transverse component of magnetic field, $\langle B_y-B_0\rangle _{\rm x,avg}/B_0$,
and the ion-density, $\langle n_{\rm i}-n_0 \rangle _{\rm x,avg}/n_0$, averaged over the $x$-domain in the 2D periodic-box simulation for the LM3.0$\beta50$ model, plotted along $\mathbf{B}_0$ ($z$-direction)
at three different times.
(c): Variations in the longitudinal component of magnetic field, $\langle B_y-B_0\rangle_{\rm x,avg}/B_0$ (red), and the ion-density, $\langle n_{\rm i}-n_0\rangle _{\rm x,avg}/n_0$ (black), averaged along the $x$-direction over the shock transition zone in the 2D shock simulation for the M3.0$\beta50$ model in HKRK2021, plotted along the $y$-direction.
Note that the preshock magnetic field, $\mathbf{B}_0^{\rm up}$, lies in the $x$-$y$ plane, and the obliquity angle between $\mathbf{B}_0^{\rm up}$ and the $y$-axis 
is $\theta_{\rm Bn}=63^{\circ}$ in the shock simulation.
\label{fig:f7}}
\end{figure}

Panel (b) of Figure \ref{fig:f7} displays the variations of the ion density, [$\langle n_{\rm i}-n_0 \rangle _{\rm x,avg}/n_0]\approx \pm 0.01$, averaged over the $x$-direction in the periodic-box simulation for the LM3.0$\beta50$ model.
Panel (a) shows the fluctuations of the transverse component of $\mathbf{B}_{0}$, $[\langle B_y-B_0\rangle_{\rm x,avg}/B_0]\approx \pm 0.4$, which have a relatively large amplitude due to the AIC-driven waves.
It shows that even after the AIC-driven waves have fully grown, they have little effects on the ion density fluctuations.

However, the ion density fluctuations of the rippling waves propagating along the shock surface behind the shock ramp are rather significant
in the shock simulation for the M3.0$\beta$50 model in HKRK2021.
Panel (c) shows that both the variations of $[\langle n_{\rm i}-n_0\rangle _{\rm x,avg}/n_0]\approx \pm 0.1$ 
and  $[\langle B_y-B_0\rangle_{\rm x,avg}/B_0]\approx \pm 0.1$ have similar amplitudes; the fluctuations of $n_{\rm i}$ are much
larger than those of the linear prediction expected for the parallel-propagating AIC mode.
Note that here the quantities are averaged along the $x$-direction over the shock transition zone including the first and second overshoot oscillations 
behind the ramp. 
Hence, the basic assumptions of the linear theory, such as the homogeneous background,
charge neutrality, and zero net-current, are likely to be violated in this region.

We point that such large-amplitude fluctuations of $n_{\rm i}$, comparable to the fluctuations of $B_{y}$, were previously recognized in the 2D hybrid 
simulations of supercritical, perpendicular shocks presented in \citet{winske1988}. 
The authors suggested that the large compressive waves might result from nonlinear effects in addition to oblique i-mirror modes.
The effects due to nonlinear couplings between various wave modes could be significant as well
\citep[e.g.][]{shukla1985,verscharen2011,marsch2011}.
Therefore, the pure AIC-driven waves in the shock transition could have been modified by such possible nonlinearities, leading to the enhancement of ion density fluctuations.

\section{Summary}
\label{sec:s5}

In supercritical $Q_{\perp}$-shocks, a substantial fraction of incoming ions and electrons are 
reflected, and the transverse components of magnetic fields are amplified at the shock ramp.
The reflected-gyrating ions and the amplified magnetic fields induce the ion and electron perpendicular temperature anisotropies,
$\mathcal{A}_{\rm i}$ and $\mathcal{A}_{\rm e}$, respectively, in the shock transition region \citep{guo2017}.
They in turn trigger various microinstabilities
including the AIC, whistler, i-mirror, and e-mirror instabilities \citep[e.g.,][]{winske1988,lowe2003,guo2017}.
The kinetic properties of these four instabilities are summarized in Table \ref{tab:t1}.
The multi-scale plasma waves generated by these microinstabilities are thought to be crucial for the electron preacceleration 
via the SSDA \citep[e.g.,][]{katou2019,niemiec2019,trotta2019}. 
 
In this work, 
adopting the numerical estimates for the ion and electron temperature anisotropies found in the 2D PIC simulations of $Q_{\perp}$-shocks with $M_{\rm s}=2-3$ (see Table \ref{tab:t2}),
we have carried out the kinetic linear analysis of the microinstabilities for wide ranges of parameters, $\beta=1-100$ and $m_i/m_e=50-1836$. 
The linear predictions for the fastest growing mode, $\gamma_m$, $k_m$, $\theta_m$, of each instability 
are given in Table \ref{tab:t2}.
In addition, in order to investigate the development and nonlinear evolution of the waves induced by the microinstabilities, 
we have performed 2D PIC simulations with periodic boundary conditions
for the three fiducial models, LM2.0$\beta50$, LM2.3$\beta50$, and LM3.0$\beta50$.
Finally, the results have been also compared with the 2D PIC simulations for ICM shocks reported in HKRK2021.

The main results can be summarized as follows:\hfill\break
1. In the LM3.0$\beta50$-m1836 model with the real mass ratio, which represents a typical supercritical ICM shock,
the maximum growth rates of the four instabilities have the following order: $\gamma_{\rm WI} \gg \gamma_{\rm EM} \gg \gamma_{\rm AIC} > \gamma_{\rm IM}$ (Fig. \ref{fig:f2}).
Hence, the parallel-propagating AIC and whistler waves are expected to be more dominant than the oblique-propagating mirror waves. \hfill\break
2. In the LM2.0$\beta50$ model, which represents a subcritical ICM shock, by contrast, the AIC mode is stable (Table \ref{tab:t2}),
 so mainly the electron-scale whistler waves are generated. \hfill\break
3. The maximum normalized growth rates for the AIC and e-mirror instabilities, $\gamma_{\rm AIC}/\Omega_{\rm ci}$ and $\gamma_{\rm EM}/\Omega_{\rm ce}$, are almost independent of $m_{\rm i}/m_{\rm e}$, while $\gamma_{\rm WI}/\Omega_{\rm ce}$ for the whistler instability and $\gamma_{\rm IM}/\Omega_{\rm ci}$ for the i-mirror instability are slightly lower for smaller $m_i/m_e$
(Fig. \ref{fig:f3}).\hfill\break
4. For all the four instabilities, the maximum normalized growth rates increase with increasing $\beta$ (Fig. \ref{fig:f3}).\hfill\break
5. As the sonic Mach number $M_{\rm s}$ increases, both $\mathcal{A}_{\rm e}$ and $\mathcal{A}_{\rm i}$ increase, 
all the modes grow faster, and $k_m$ of each mode shifts towards larger $k$, 
regardless of $\beta$ in the range of $\beta\approx 20-100$ (Fig. \ref{fig:f4}).\hfill\break
6. The critical sonic Mach number to trigger the AIC instability in the shock transition 
is $M_{\rm AIC}^{*}\sim 2.3$ for $\beta\approx 20-100$. 
It is similar to the EFI critical number suggested in our previous study,
that is, $M_{\rm AIC}^{*} \approx M_{\rm ef}^{*} \approx 2.3$
\citep{kang2019}. 
For $\beta=1$, on the other hand, $M_{\rm AIC}^{*}\gtrsim 3$ is slightly higher, 
because the AIC mode is suppressed by the strong magnetization of ions \citep{hellinger1997,trotta2019}.\hfill\break
7. The 2D  periodic-box simulations confirm the linear predictions.
In the early stage of $\sim \tau_{\rm WI}$, electron-scale waves develop and then decay as $\mathcal{A}_{\rm e}$ is relaxed, followed by
the growth of ion-scale waves on the time scale of $\sim \tau_{\rm AIC}$ (Figs. \ref{fig:f5} and \ref{fig:f6}).\hfill\break
8. The rippling waves propagating along the shock surface have the characteristics of AIC waves. 
Although the AIC waves are parallel-propagating, electromagnetic, incompressible in the linear regime, the amplitudes of the longitudinal magnetic field and ion-density fluctuations associated with the overshoots in the shock transition are similar and of the order of 10\% according to the shock simulation for the M3.0$\beta50$ model (Fig. \ref{fig:f7}).
It is expected that the inhomogeneity in the shock transition and the nonlinear effects could lead to
the generation of such large-amplitude fluctuations of the ion-density along the shock surface. 
\hfill\break

In conclusion, our results well support the suggestion for the generation of multi-scale plasma waves via various microinstabilities
in the transition region of high-$ \beta$, supercritical, $Q_{\perp}$-shocks \citep{guo2017,katou2019,niemiec2019,trotta2019}. 
A detailed description of the shock structure and the electron preacceleration in such ICM shocks, 
realized in 2D PIC simulations, is reported in HKRK2021.

\acknowledgments
{\color{black}The authors thank the anonymous referee for constructive comments.}
This research used the high performance computing resources of the UNIST Supercomputing Center. 
S.K. was supported by the NRF grant funded by the Korea government (MSIT)(NRF-2020R1C1C1012112).
J.-H. H. and D.R. were supported by the National Research Foundation (NRF) of Korea through grants 2016R1A5A1013277 and 2020R1A2C2102800. 
J.-H. H. was also supported by the Global PhD Fellowship of the NRF through grant 2017H1A2A1042370.  
H.K. was supported by the Basic Science Research Program of the NRF through grant 2020R1F1A1048189.

\bibliography{ms_shock}{}
\bibliographystyle{aasjournal}

\end{document}